\documentclass[12pt]{article}
\usepackage{abstract}

\usepackage{booktabs}
\usepackage{mathastext}
\usepackage{times}
\usepackage{amsmath}
\usepackage{graphicx}
\usepackage{float}
\usepackage{amssymb}

\makeatletter
\newcommand*\redefscriptstyle{\mathpalette\redef@scriptstyle\relax}
\newcommand*\redef@scriptstyle[2]{\everymath{%
  \ifx\protect\@typeset@protect
    \expandafter\expandafter\expandafter\redef@finish
  \fi
  \scriptstyle\redef@setsf}}
\newcommand*\redef@setsf{\mathsf{\xdef\protect@sf{\mathgroup\the\mathgroup\relax}}}
\newcommand*\redef@finish[1]{%
  \let\protect\@typeset@protect
  \redef@setsf}
\makeatother

\usepackage{bm}

\let\oldbm\bm
\renewcommand{\bm}[1]{\boldsymbol{\oldbm{#1}}}

\title{Highly sensitive single-molecule detection \\in slow protein ion beams} 
\author{
  M. Strau\ss$^{1}$,
  A. Shayeghi$^{1}$,
  M.F.X. Mauser$^{1}$,
  P. Geyer$^{1}$, \\
  T. Kostersitz$^{1}$,
  J. Salapa$^{1}$,
  O. Dobrovolskiy$^{1}$,
  S. Daly$^{2}$,
  J. Commandeur$^{2}$, \\
  Y. Hua$^{3}$,
  V. Köhler$^{3}$,
  M. Mayor$^{3}$,
  J. Benserhir$^{4}$,
  C. Bruschini$^{4}$, \\
  E. Charbon$^{4}$,
  M. Castaneda$^{5}$,
  M. Gevers$^{5}$,
  R. Gourgues$^{5}$,
  N. Kalhor$^{5}$, \\
  A. Fognini$^{5}$,
  and M. Arndt$^{1\ast}$
  \\[6pt]
  \footnotesize $^{1}$University of Vienna, Faculty of Physics \& VDSP \& VCQ, Boltzmanngasse 5, A-1090 Vienna \\
  \footnotesize $^{2}$MSVISION, Televisieweg 40, 1322 AM Almere, The Netherlands \\
  \footnotesize $^{3}$Department of Chemistry, University of Basel, St. Johannsring 19, CH-4056 Basel, Switzerland \\
  \footnotesize $^{4}$Advanced Quantum Architecture Laboratory, EPFL, Rue de la Maladière 71b, CH-2002 Neuchâtel, Switzerland \\
  \footnotesize $^{5}$Single Quantum, Rotterdamseweg 394, 2629 HH, Delft, The Netherlands \\
  \footnotesize $^\ast$Correspondence should be sent to: markus.arndt@univie.ac.at
}
\date{\footnotesize June 26, 2023}

\begin{document} 
\maketitle

\begin{abstract}
The analysis of proteins in the gas phase benefits from detectors that exhibit high efficiency and precise spatial resolution. Although modern secondary electron multipliers already address numerous analytical requirements, new methods are desired for macromolecules at low energy. Previous studies have proven the sensitivity of superconducting detectors to high-energy particles in time-of-flight mass spectrometry. Here we explore a new energy regime and demonstrate that superconducting nanowire detectors are exceptionally well suited for quadrupole mass spectrometry. Our detectors exhibit an outstanding quantum yield at remarkably low impact energies. Notably, at low ion energy, their sensitivity surpasses conventional ion detectors by three orders of magnitude, and they offer the possibility to discriminate molecules by their impact energy and charge. By combining these detectors into arrays, we demonstrate low-energy ion beam profilometry, while our cryogenic electronics pave the way for future developments of highly integrated detectors.
\end{abstract} 

\newpage
\section*{Introduction}
Mass spectrometry is a ubiquitous tool in the life sciences, chemistry, and physics, allowing detection, identification, and analysis of objects ranging in size from atoms to large biopolymers. Some of the most wide-spread instruments, such as quadrupole (QMS) \cite{Paul1990} and time-of-flight mass spectrometers (TOF-MS), use detectors that rely on secondary electron multiplication (SEM). While SEM detectors have a quantum yield of $\eta \simeq$ 90 \% for electrons \cite{Fehre2018},  the efficiency for macromolecules depends strongly on their velocity, mass, and structure \cite{Brunelle1997,Beuhler1980} and can drop to $\eta \simeq 10^{-5}$ for proteins with kinetic energies of only a few tens of electron volts.
It only becomes sizeable for impact velocities of $v > 20$\,km/s, i.e. an impact energy of $40$\,keV for masses beyond $20$\,kDa \cite{Twerenbold2001}. Since more than $80\,$\% of all known proteins are found in this mass regime \cite{Berman2003}, efficient detectors are highly desirable. 

\begin{figure}[bh]
\centering
\includegraphics[width=12cm]{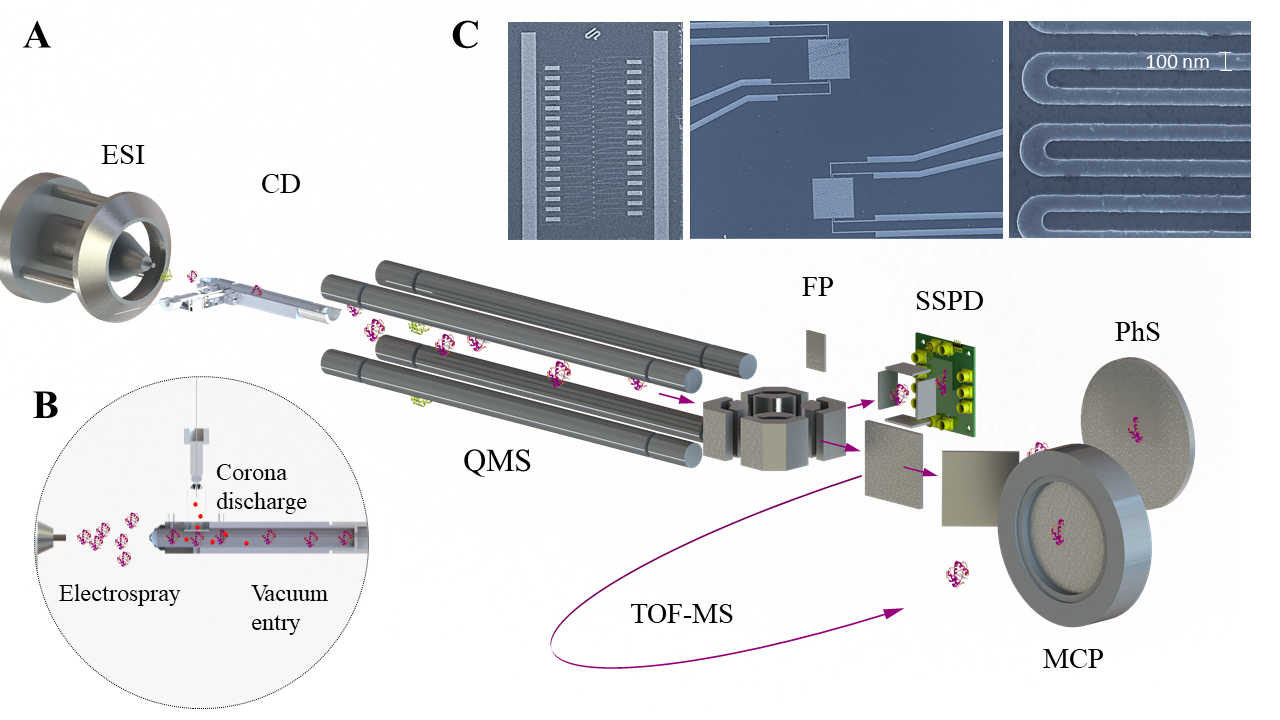}
\caption{Quadrupole mass spectrometry with quantum detection (ESI-QMS-QD). (A) Proteins are volatilized using electrospray ionization (ESI) and charge-reduced in bipolar air by a corona discharge (CD). They are filtered in a quadrupole mass selector (QMS) and fly through a radio frequency hexapole guide (not shown) towards a quadrupole bender. This steers them to either a time-of-flight mass spectrometer (TOF-MS), a phosphor screen (PhS), or the superconducting single particle detector array (SSPD). A Faraday plate (FP) can be shifted into the position of the SSPD to calibrate the incident ion current in the limit of high flux. (B) Corona charge reduction of high mass proteins (C) Multi-pixel array and meander structure $D_1$.}
\end{figure}

The detection challenge is alleviated when working with highly charged ions since their kinetic energy is proportional to their charge q and the accelerating electric potential $U_{acc}$. On the other hand, a low charge state is often favorable as it reduces the complexity of the mass spectrum and facilitates the identification of spectra lines. In that case detectors are required which can efficiently detect slow ions. The capability to analyze biomolecules at low velocity and long interaction time can also open additional new analytic opportunities in combination with optical spectroscopy and deflectometry. Here, we explore superconducting nanowires as quantum detectors for biomolecules at energies about 100 times smaller than commonly required for SEM detection, providing the same or even better spatial resolution, with a potential to improve on that by more than an order of magnitude.	

Superconducting detectors are intriguing as they possess a small energy gap in the few meV regime that allows them to be highly sensitive to low energy quanta. This gap can be bridged by the impact energy of a particle or by the absorption of a photon. A variety of cryogenic sensors have found interdisciplinary application in numerous research fields in form of bolometers \cite{Hilton1998}, transition edge sensors \cite{Irwin1995,Irwin2005}, kinetic inductance detectors \cite{Kerman2006}, superconducting tunneling junctions \cite{Wood1969,Twerenbold1996}, or superconducting nanowire detectors (SNWDs), also known as superconducting single-particle detectors (SSPDs) \cite{Semenov2001,Ohkubo2008}. All these sensors found important use cases in photonics and a few of them have also been successfully used in TOF-MS \cite{Twerenbold1996}, among them transition edge sensors \cite{Irwin1995}, superconducting tunnel junctions and SSPDs.	

In comparison to superconducting tunnel junctions, SSPDs combine good spatial and temporal resolution with a working temperature above 3 K, where substantially higher cooling power is available at moderate cost \cite{Cristiano2015}. This is important for up-scaling current prototypes to multi-pixel devices, which are needed for most applications. SSPDs are typically etched into a superconducting film in a meander structure on top of a silicon-based substrate. The nanowire is driven by a bias current $I_b$ close to its critical current, typically in the 10 - 100 µA range. If a particle impacts the detector, the energy released to the superconductor locally breaks up Cooper pairs, causing a quantum-phase transition from the superconducting to the normal conducting state. As a result, a normal conducting hot spot is generated. When the current around the resistive area exceeds the local critical current density, the strip will become normal conducting along its full width. The continuous driving of a current into the now resistive strip leads to a fast voltage peak that triggers the signal. Thermal relaxation to the substrate resets the detector. 

SSPDs were first developed for applications in photonics \cite{Semenov2001} where response times below 20 ps \cite{Pernice2012,Korzh2020}  are possible. Detection efficiencies of up to 99 \% can be realized even at telecom wavelengths \cite{Zadeh2021}. In addition, single photon sensitivity at MIR range of 10 µm wavelengths have been demonstrated \cite{Verma2021}.  SSPDs are impact detectors but were also designed to have photon number resolution \cite{Divochiy2008}. They have found applications in the explorations of the foundations of physics \cite{Giustina2015, Shalm2015}, in quantum optics \cite{Takesue2007,Aker2022} and quantum information processing \cite{Zadeh2021,Mirhosseini2020}, astronomy \cite{Wollman2019} and molecular science \cite{Chen2017}. Recently, nanowire detectors have also been used in TOF-MS at high kinetic energy \cite{Ohkubo2008, Cristiano2015,Suzuki2008a,Suzuki2008b}, always for the explorations of fast molecular beams. However, in the low-energy regime only a single study has been reported, without protein mass identification \cite{Marksteiner2009}. 

Here, we explore high-resolution quadrupole mass spectrometry of proteins in low charge states, with highly efficient superconducting nanowire detector arrays, at energies well below the requirements of conventional secondary electron detectors. To upscale this technology, we study different detector sizes and geometries and discuss how low-noise cryogenic electronics can enable large multi-pixel arrays.

Our detectors are manufactured as arrays of 10 nm thick NbTiN nanowire meanders, each of which with a filling factor of 50 \% and a critical temperature of $T_c \simeq$ 10 K \cite{Gourgues2019}. A pulse tube cooler keeps the SSPD at an operating temperature of 3.7 K. 	  
Detector $D_1$, fills an area $A_1$ of 20 $\times$ 20 $\mu m^2$ in meander form with a line width $w_1$  of 100 nm and a pitch $p_1$ of 200 nm. These small detectors are highly sensitive, fast, and spatially resolving at low particle energies. In order to increase the effective detection area, a 32-pixel linear array was fabricated, where all active pixels were connected via individual leads to the signal conditioning and processing electronics at room temperature. Future parallelization to hundreds of elements will require cryogenic onboard electronics, which we here explore as chip $D_{1,b}$, augmented by one low-noise amplifier (LNA) per pixel. 	
The purpose of chipset $D_1$ is to realize a multi pixel array with high spatial and temporal resolution. In molecular beam research and mass spectrometry, however, total detector size is also important. Therefore, the pixel area was increased to  $A_2$ = 200 $\times$ 200 $\mu m^2$, yielding a hundred-fold detection area increase. Detector $D_2$ combines ten such pixels, each with a single line with $w_2$ = 500 nm and $p_2$  = 1000 nm. 

\section*{Results}

\subsection*{Protein Mass Spectra}
A tandem mass spectrometer was extended by a cryogenic vacuum chamber which hosts the nanowire detectors (Figure 1). Proteins are ionized using electrospray ionization \cite{Fenn1989} and optionally charge reduced by corona discharge \cite{Ebeling2000}. The analyte ions are guided into a customized quadrupole mass selector, which can be scanned with a resolution of $\Delta m/m \simeq 1:1000$ up to a mass to charge ratio of m/z = 30 000 Th. A hexapole ion guide and subsequent ion deflector steer the ions either towards a conventional TOF-MS with MCP detector, or a phosphor screen (PhS) with a counting photo-multiplier. Additionally, the ions can be directed towards the SSPD with the use of a quadrupole deflector. The MCP area $A_{MCP}$= 700 $mm^2$ is about 100 times larger than the ion beam and 5000 times larger than a single pixel of detector $D_2$. We have therefore chosen an integration time of 120 s for the classical channel and 1200 s for the sum of six quantum channels to achieve comparable signal-to-noise levels.
\begin{figure}[ht]
\centering
\includegraphics[height=7cm]{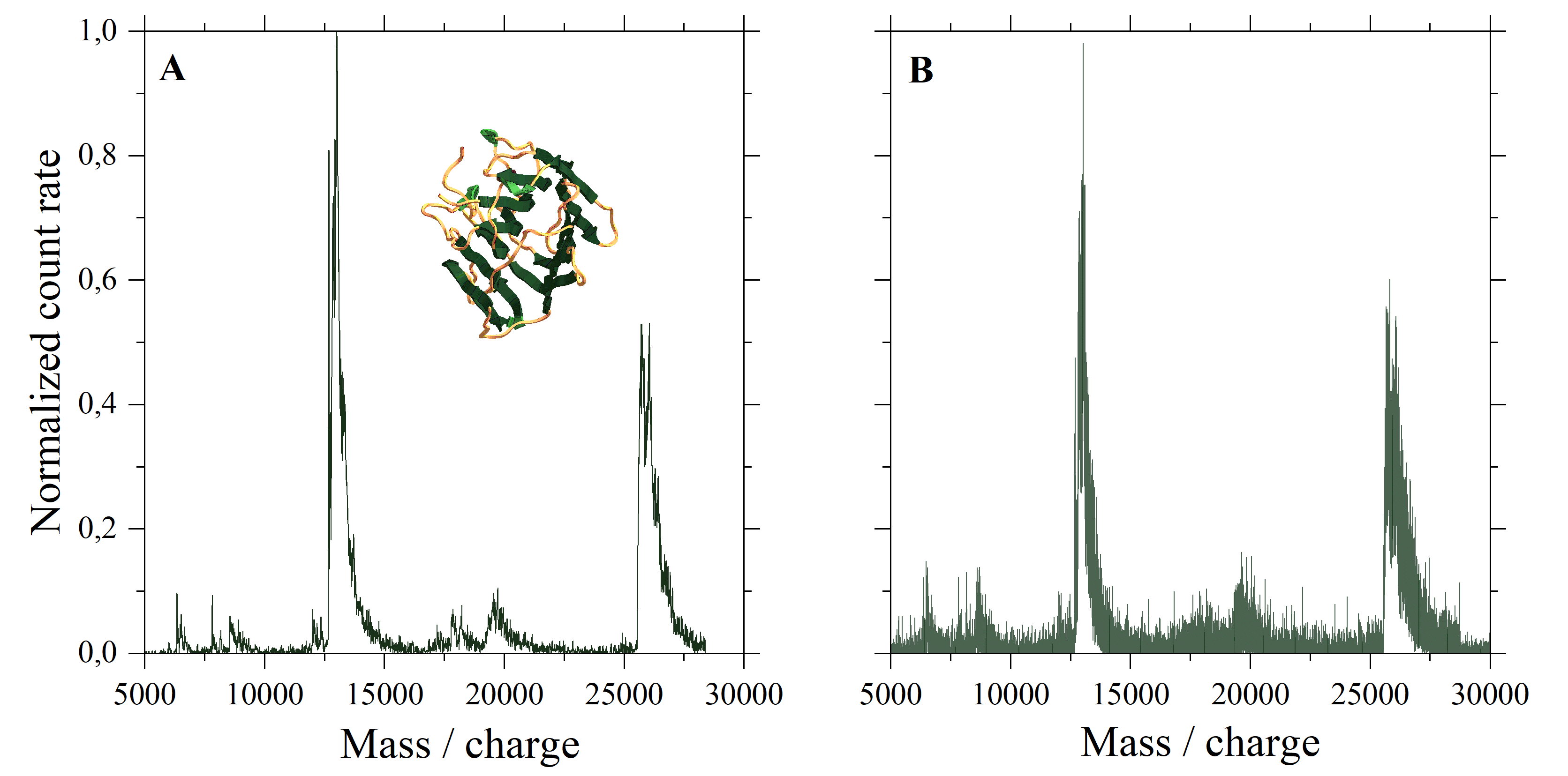}
\caption{Quantum and classical detection of an ESI-quadrupole mass spectrum of high mass proteins.  Charge-reduced concanavaline A (27 kDa) is detected by (A) the quantum SSPD array $D_2$ at $U_{acc}$ = 190 V  (B) the classical phosphor screen with $U_{acc}$ = 10 kV.  }
\end{figure}

The kinetic energy of the ions at each detector, $E_{kin} = q U_{acc}$, is determined by their charge q and by the potential difference $U_{acc}$ of the detector with respect to the last acceleration electrode. In TOF-MS, the potential  $U_{acc}$ reaches 13 kV, while on the phosphor screen it is limited to $U_{acc} $ =  10 kV. In contrast, we operate our quantum detector at $U_{acc}\simeq$ 0.1 – 0.2 kV and find that it can accept 100 times lower impact energies than required by conventional devices. We could determine the detection efficiency quantitatively for impact energies down to 100 eV and can observe well resolved mass spectra down to 40 eV. We demonstrate this in Figure 2, where we present the QMS selected mass spectrum of charge reduced concanavaline A (27 kDa). The signal to noise ratio (SNR) of the SSPD detector competes favorably with the conventional phosphor screen and MCP, even for a 50 times lower acceleration voltage.

\begin{figure}[ht]
\centering
\includegraphics[height=12cm]{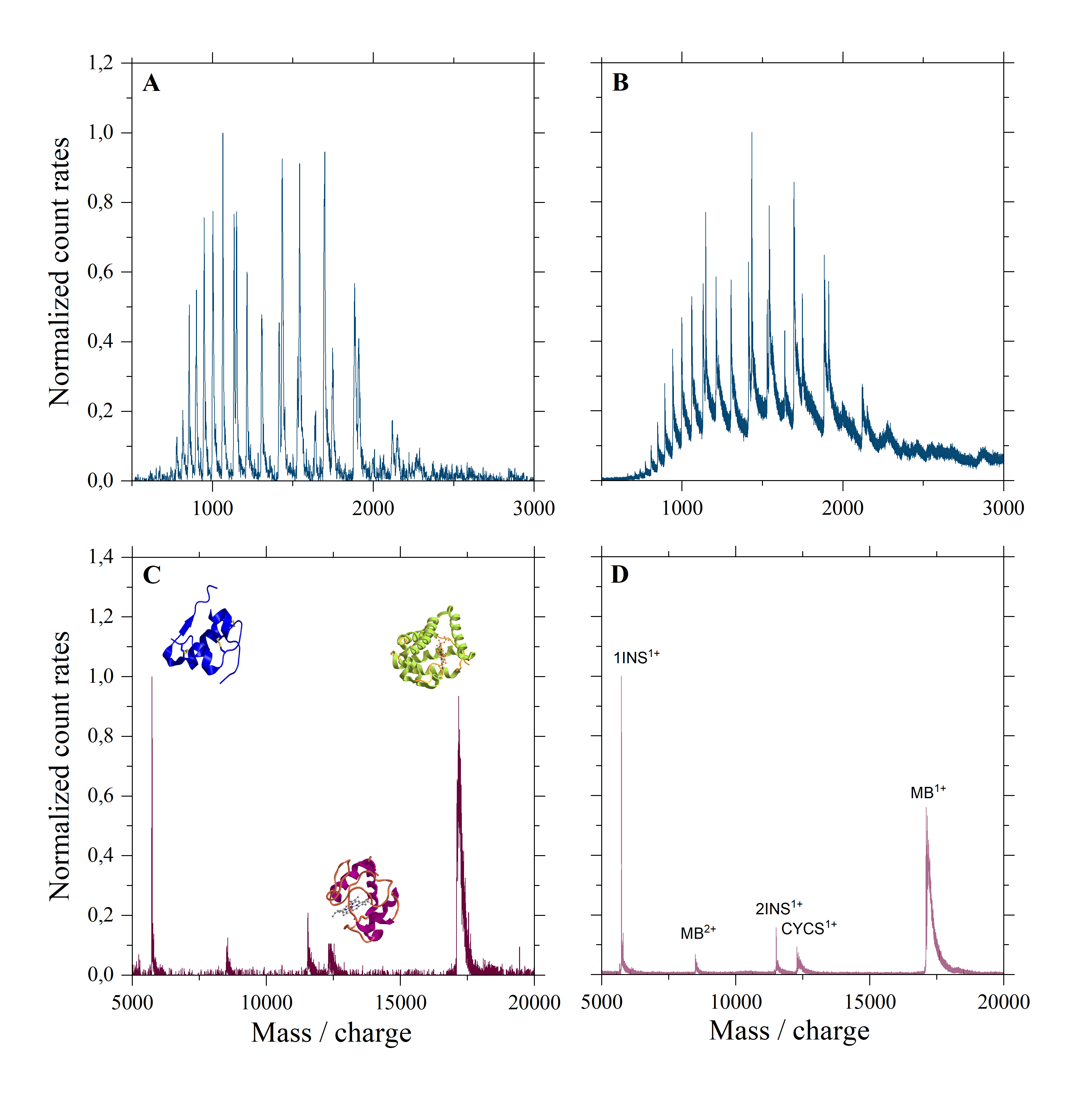}
\caption{Comparison of protein mass spectra in ESI-QMS-SSPD an ESI-TOF-MCP. Insulin (5.7 kDa), cytochrome C (12.3 kDa) and myoglobin (17 kDa) cations are electrosprayed and analyzed by a quadrupole mass filter with SSPD detector (A, C) and by a time-of-flight mass spectrometer with an MCP detector (B, D). The dense group of peaks seen in sub-figure A and B is associated with multiply charged proteins without any charge reduction (see Suppl. Inf.), while the lower spectra C and D are enormously simplified by charge-reduction. Here, post-acceleration occurs at an acceleration voltage of $U_{acc}$ = 190 V and $U_{acc}$ =13 000 V for the SSPD and the MCP, respectively.}
\end{figure}

To demonstrate high mass and low energy detection of various compounds in a wide mass range, we show in Figure 3 (A, C) a protein mass spectrum obtained by high-resolution quadrupole mass spectrometry combined with low energy SSPD detection and compare it with a TOF-MS measured with an MCP detector in Figure 3 (B, D). For this we analyze a protein mix of insulin (5.7 kDa), cytochrome C (12.3 kDa) and myoglobin (17 kDa), after electrospray  ionization Figure 3  (A, B) and charge reduction Figure 3 (C, D). The charge reduction drastically simplifies the spectral assignment and profits from the high sensitivity of the SSPD. Note that here the SSPD achieves a similar quantum yield per detector area as the MCP. However, it does this at ion energy, where conventional SEM detectors are 1000 times less efficient.

\subsection*{Energy and charge dependence}
Having proven SSPDs as excellent molecular impact counters for QMS-MS, we now analyze, how they can additionally contribute to assessing the charge and impact energy of an ion, a feature exceeding the capability of most mass spectrometers. In Figure 4 we plot the insulin ion count rate as a function of the bias current $I_b$ for a single pixel of $D_2$. The curves are taken for varying charge states q and different values of the acceleration voltage $U_{acc}$. We find that ions of the same kinetic energy $E_{kin}$ fall onto the same curve, while ions at different kinetic energies (380 eV, 190 eV, 95 eV) can be clearly discerned by the onset of the signal as a function of the bias current.

\begin{figure}[ht]
\centering
\includegraphics[height=7cm]{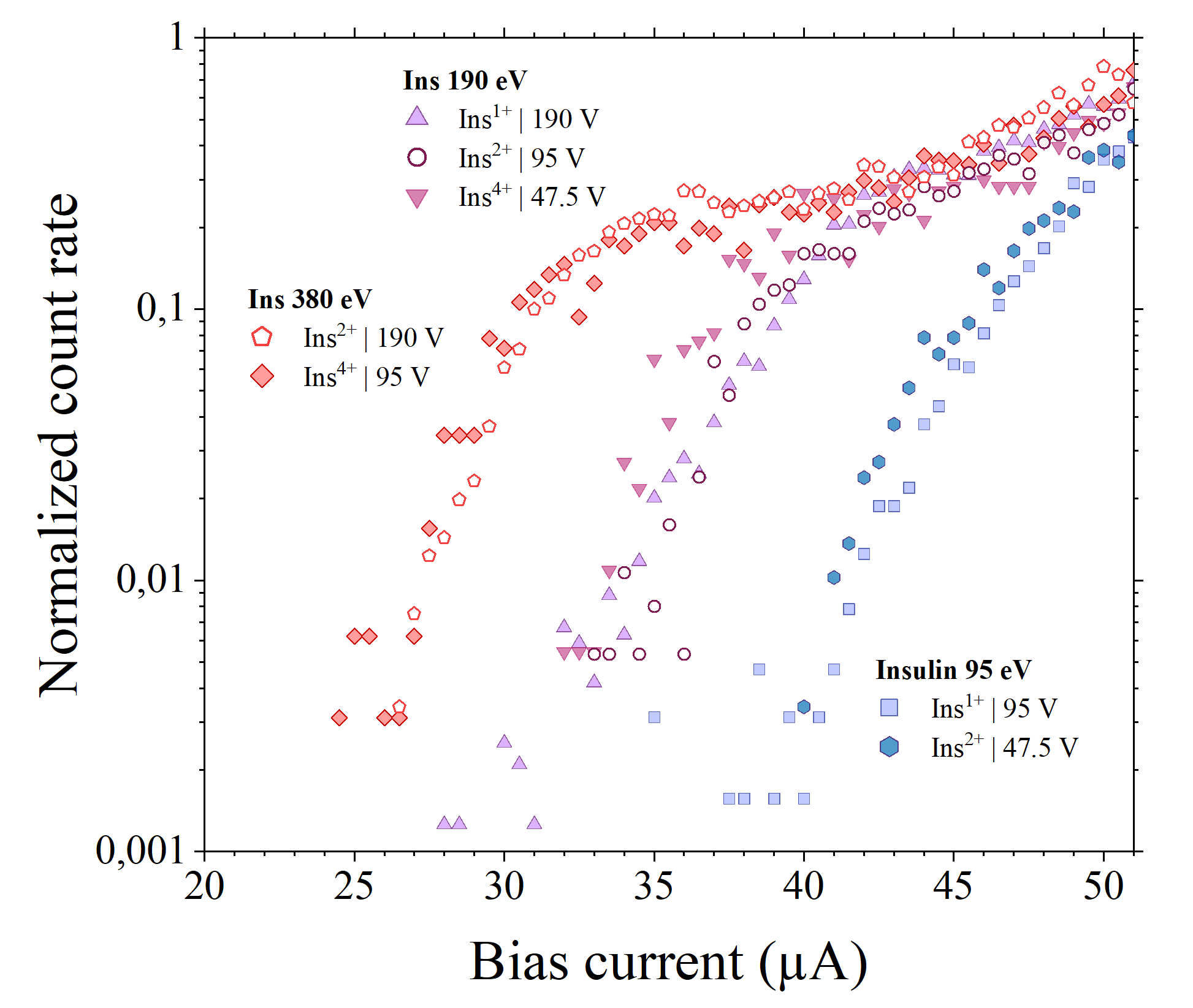}
\caption{Energy-dependence of the SSPD detection efficiency. The normalized protein count rate is shown as a function of the SSPD bias current $I_b$, for insulin in three  different charge states (q = +1, +2, +4) and at three different acceleration voltages ($U_{acc}$ = 47.5, 95 V, 190 V). Three sets of curves confirm that the normalized count rate clearly depends on the kinetic energy $E_{kin}=q~U_{acc}$, but not separately on charge.}
\end{figure}

These curves suggest that under our experimental conditions an energy resolution of 20 to 30 eV can be achieved. Our result also shows that ions of the same mass-to-charge ratio but different charge, such as for instance a doubly charged insulin dimer and a singly charged insulin monomer will be distinguishable when monitoring the SSPD bias current curves in addition to the QMS transmission. A better energy resolution will be achieved in future device designs.

\subsection*{Confirming the hot-spot model}
In all bias current curves shown in Figure 4 we can determine a threshold bias current $I_{th}$ via the intersection of the asymptotes at low and high bias current \cite{Verevkin2002}. This is shown in Figure 5A for singly charged insulin. 
The dependence of $I_{th}$ on the molecular impact energy, is shown in Figure 5B. 
It is well reproduced by the analytical form $I_{th} = I_c (1- \gamma \sqrt{E_{kin}}/ w_2)$,  which is predicted by a normal-core hot-spot model \cite{Suzuki2011}. 
From a fit to the data, we can deduce the critical current $I_c$ = 58.1 µA, from which we learn about the underlying physical transduction mechanism by comparing $I_c$  with the depairing current  
$I_{dep}(T)= I_{dep}(0) \left[ 1-(T/T_c)^2 \right]^{3/2}$. 

Here $I_{\text{dep}} = 0.74 w \Delta^{3/2}(0)/(e R_{\square} \hbar D)$, where $\Delta(0)$ is the superconducting energy gap at $T = 0\,\text{K}$, $e$ is the electron charge, $R_{\square}$ is the sheet resistance, and $D$ is the electron diffusion coefficient \cite{Bezuglyj2022}.
With $R_{\square}$ = 380 $\Omega$, $T_c$ = 10.5 K, and D = 0.48 $cm^2/s$ for NbTiN \cite{Sidorova2021}, we obtain $I_{dep} (0) \simeq 350 \mu$A and $I_{dep} (3.7 K) \simeq 285$ $\mu A$. 

\begin{figure}[ht]
\centering
\includegraphics[height=7cm]{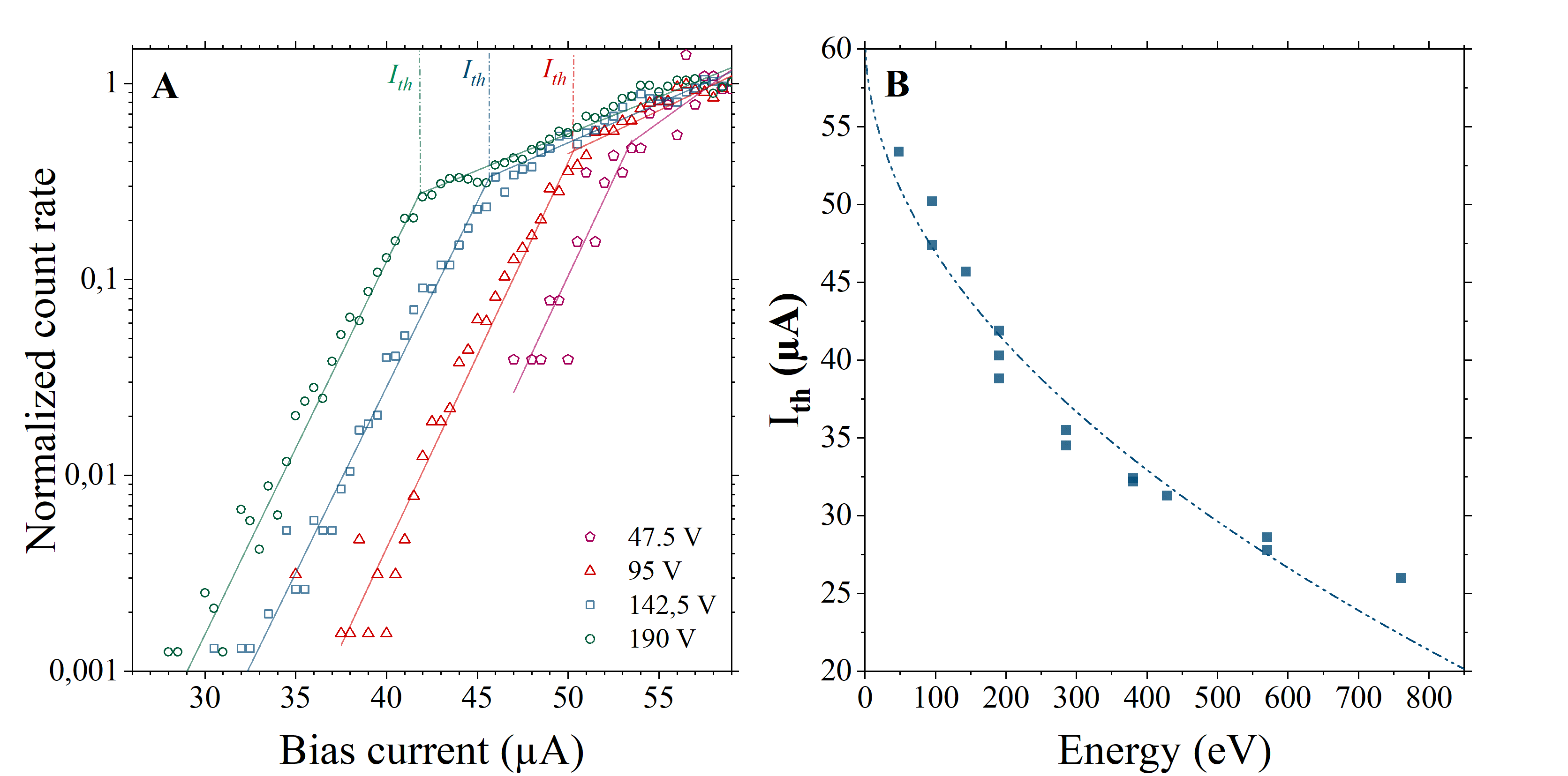}
\caption{Corroboration of the hot spot detection model for proteins. Dependence of the threshold bias current $I_{th}$ on the impact energy. (A) $I_{th}$ is found at the intersection of the two asymptotes (straight lines) of the normalized count rate plotted as a function of bias current $I_b$. We show data for singly charged insulin impacting on a single pixel of detector $D_2$. (B) $I_{th}$ as a function of the kinetic energy $E_{kin} =q~U_{acc}$. The dashed line represents the analytical normal-core hot-spot model, which fits the experiment very well.}
\end{figure}

Since the critical current amounts to only 20 \% of the depairing current $I_c/I_{dep}$ = 0.2, this supports a hot spot detection mechanism \cite{Ohkubo2008} and excludes an alternative vortex-assisted detection model \cite{Vodolazov2017}, as this would require $I_c/I_{dep} > 0.5$. The hot-spot diameter in NbTiN is here estimated to be $D_{HS} = 4\sqrt{\tau~t~h~D} \simeq$ 35 nm, where $\tau_{th}$ is the thermalization time. We conclude that detector $D_2$ is working in the hotspot regime. 

\subsection*{Momentum dependence}
For particles to create a normal-conducting hotspot, they must couple to either the charge carriers or the phonons in the nanowire. It is therefore interesting to explore if it is possible to distinguish between the two effects. Since for fixed kinetic energy the momentum transfer increases with mass, $p = (2mE_{kin})^{1/2}$, we study the bias current curves for molecules of the same kinetic energy, $E_{kin}$ = 190 eV, but different mass, momentum and number of constituents. Figure 6 shows count rates of detector type $D_2$ for singly charged rhodamine 6G (C$_{28}$H$_{31}$N$_{2}$O$_{3}$Cl, $N_{atoms} = 65$, MW $= 479\,$Da), bovine insulin (C$_{254}$H$_{377}$N$_{65}$O$_{75}$S$_{6}$, $N_{atoms}$ = 777, MW = 5733 Da) and myoglobin (153 AA residues, $N_{atoms}$ = 1304, MW = 16952 Da). They cover a mass ratio of about 1:12:35, a momentum ratio of about 1:3.5:6 and a ratio of atom numbers of about 1:12:20.

\begin{figure}[ht]
\centering
\includegraphics[height=7cm]{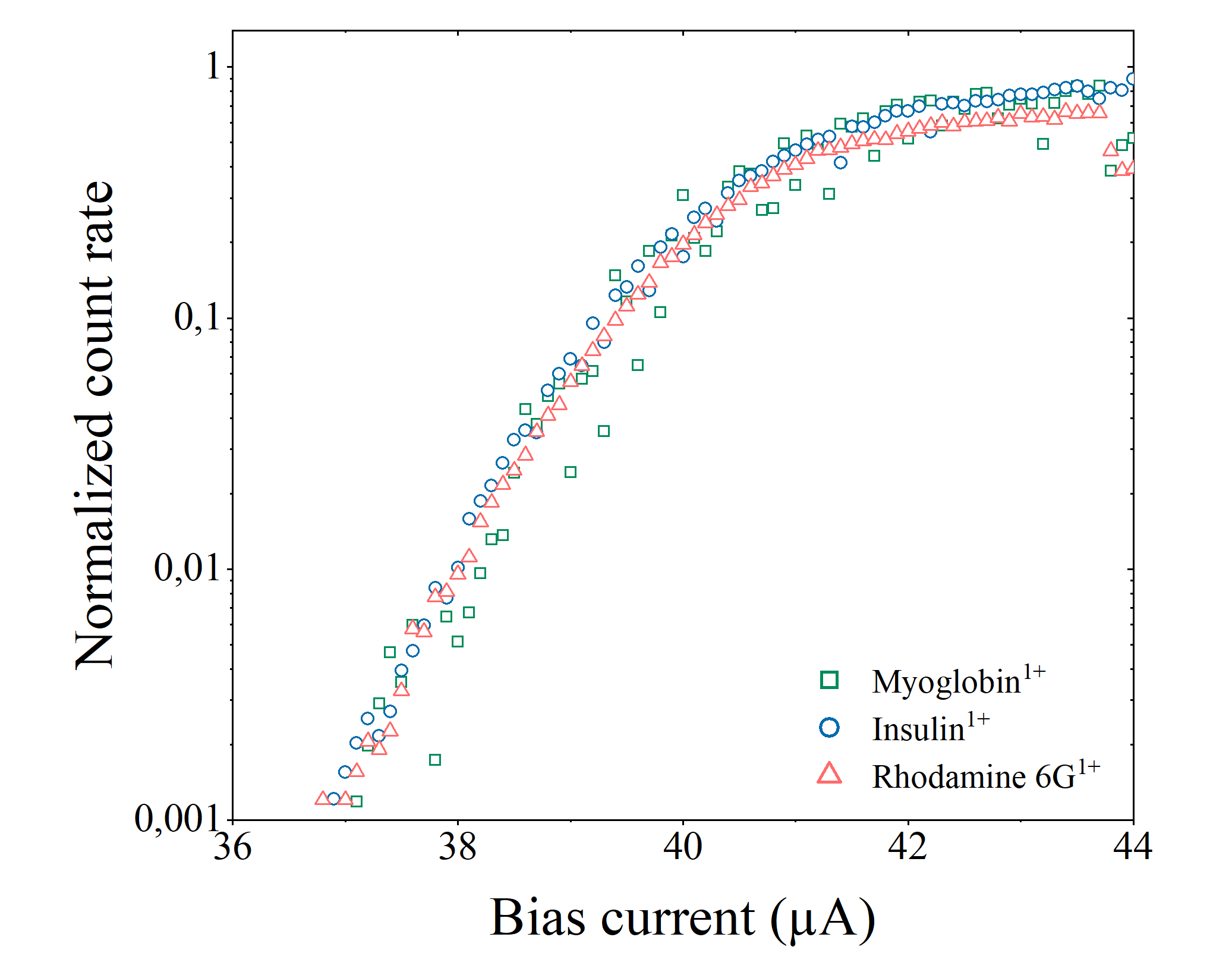}
\caption{Mass/momentum/structure-independence of the SSPD detection efficiency. At equal impact energy (190 eV) and charge we find the same normalized detector response curve for molecules of vastly different structure and complexity, specifically rhodamine 6G, insulin, and myoglobin. This suggests that kinetic energy is of prime relevance in the transduction mechanism, while structure, atom number, charge and mass are not. All measurements were made using a single pixel of detector array $D_2$.}
\end{figure}

Here we find that neither mass nor momentum nor structural complexity seem to play a role under these experimental conditions. The detection mechanism seems to depend only on the impact energy. This is interesting as one may expect to see the influence of competing energy loss channels, such as the deformation or dissociation of a molecule during impact, which all depend on atom number and complexity. Future experiments will explore the detection mechanism at ion energies down to 1 eV, which were not yet accessible in our present experiment. This composition and mass-independence is a plus and we use it as a feature in the calibration of the absolute detector quantum yield below. 

\subsection*{Protein ion beam profiling}
Given the high quantum efficiency of our nanowire detector for massive particles, its good time response and spatial resolution, one can record a spatial image of the protein beam. We do this using detector $D_2$ instead of $D_1$ to trade resolution for higher count rates. Figure 7 shows that this already provides well-resolved and useful images to optimize steering optics in ion beam trajectories in mass spectrometry. A pair of X/Y-electrodes in front of the SSPD detector are used to deflect the incident ion beam with high reproducibility, which can be calibrated in one axis against a mechanical translation, with 30 µm accuracy. We use it here to record a spatial profile of $insulin^{+5}$ with impact energy $E_{kin}$ = 1 keV. Note that the ion beam is asymmetric because of the cylindrical symmetry of the ion bender.

\begin{figure}[ht]
\centering
\includegraphics[height=7cm]{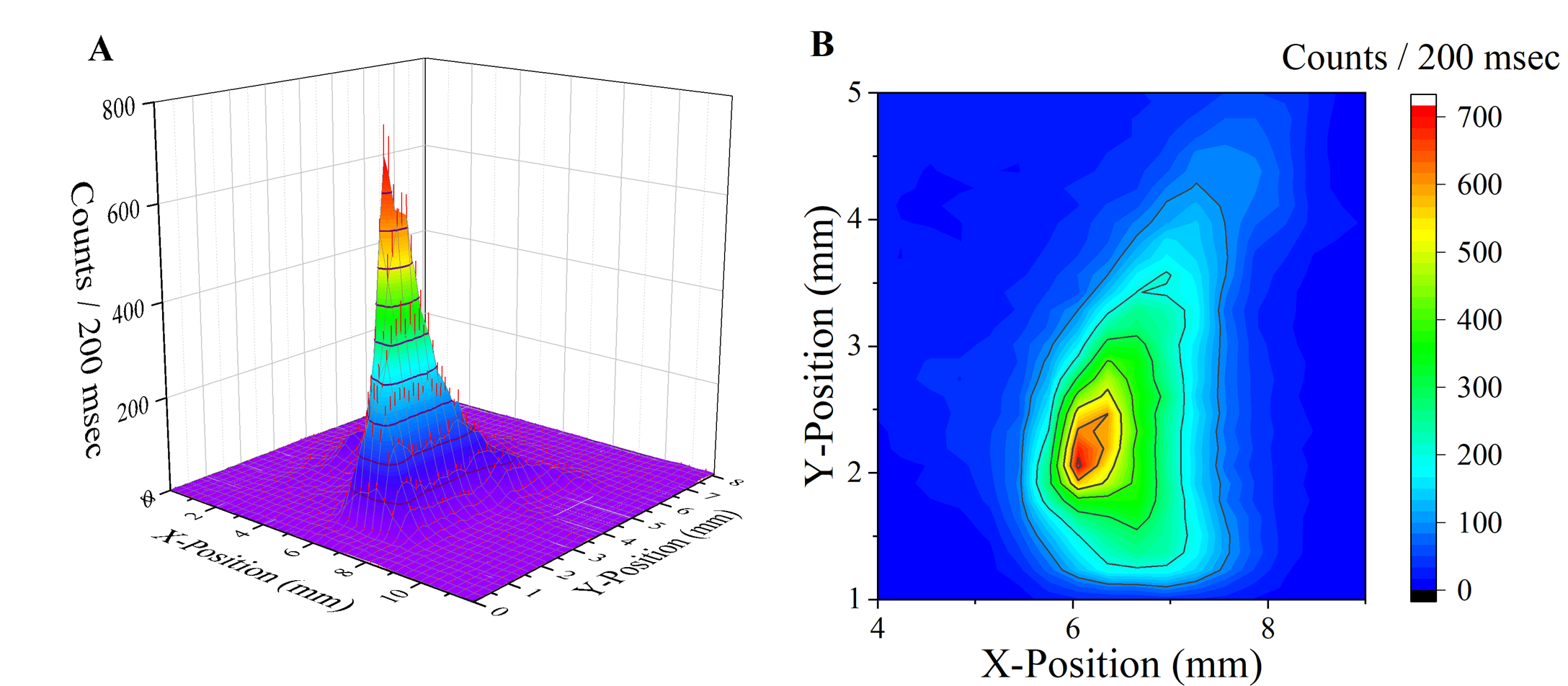}
\caption{SSPD detection of a mass-selected Insulin ion beam profile. The X position represents the mechanical shift of the SSPD array. The Y shift quantifies the ion beam deflection in an external field (see Suppl. Inf). The acquisition time for this profile was approximately 1 hour, using a single pixel of detector $D_2$.}
\end{figure}

For beams of high flux, the resolution can be improved by using detector $D_1$. However, the two SSPD detector geometries differ in absolute counts (by a factor of about 100 in pixel area) and in dark count rate. While $D_1$ is sensitive to photons and must be shielded from stray radiation, when operated at high bias current, $D_2$ is blind to black body radiation and achieves a dark count rate as low as 0.022 cps.
On the other hand, $D_1$ offers a ten-fold better spatial resolution than $D_2$, which is competitive with state-of-the-art MCPs and an advantage for molecular beam profilometry and deflectometry. 

\subsection*{Absolute detection quantum yield}
Superconducting nanowire detectors are known to be efficient photon detectors, but varying numbers have been reported in literature on the absolute quantum yield for low-energy ions \cite{Sclafani2012} and absolute calibration for molecules at low kinetic energies has been a challenge. Normalized count rates as shown in Figure 6, may lead one to assume that observing saturation is an indicator for a detection yield of $100\,$\%. However, this may be wrong for fast atomic ions, which can penetrate the nanowire and therefore deposit only a fraction of their energy. It could also be wrong for fast proteins if they converted impact energy into internal vibrational excitation or even dissociation.  

\begin{figure}[ht]
\centering
\includegraphics[height=7cm]{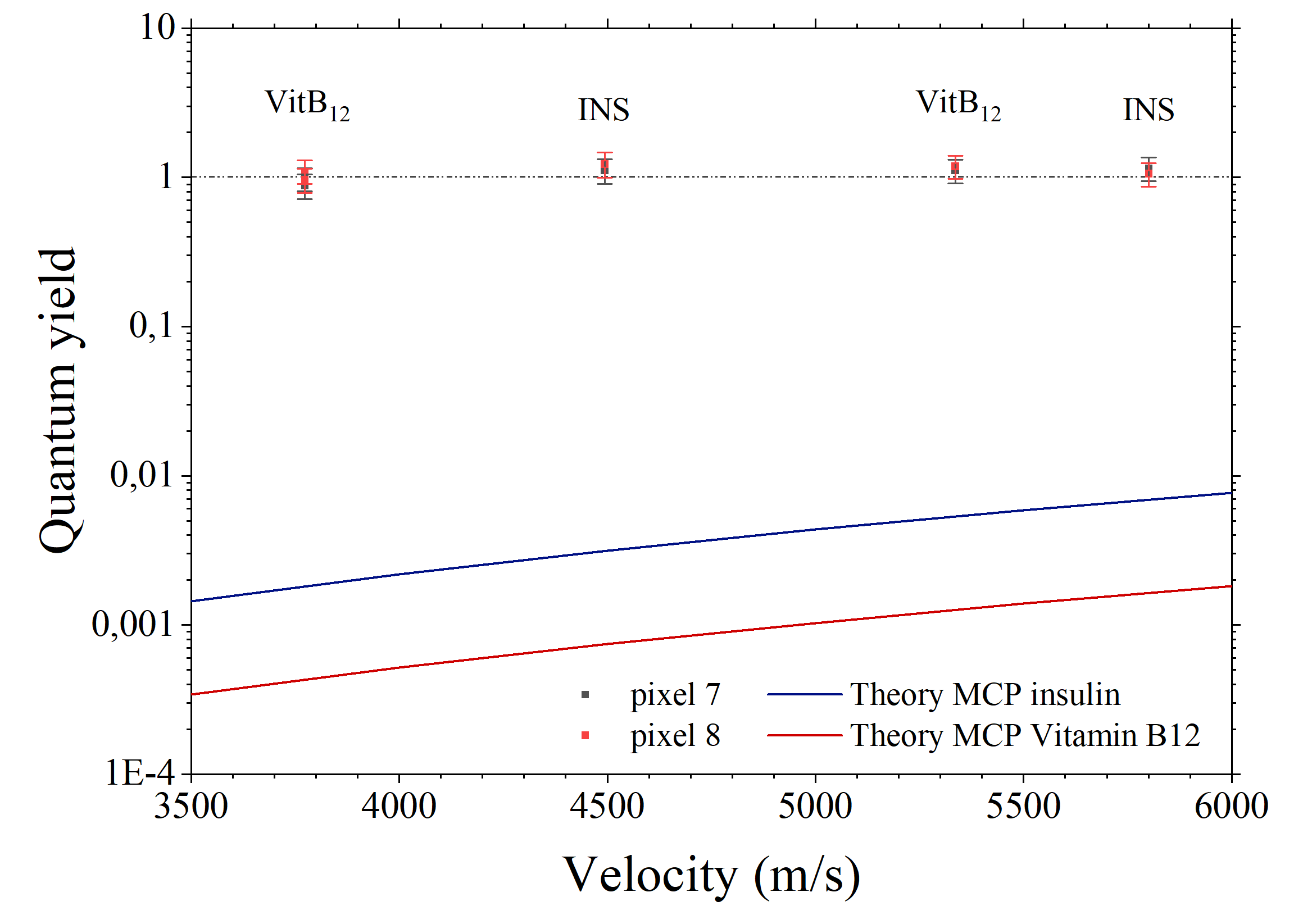}
\caption{Detection quantum yield for large molecules on a SSPD and a MCP detector. The grey and red squares show the detection quantum yield of two different pixels of detector $D_2$ as a function of the impact velocity (energy). From left to right: Vitamin $B_{12}$ at 100 eV and 200 eV, Insulin at 600 eV and 1000 eV of kinetic energy. These molecules have different atom numbers, masses, momenta, and energies, but they all have detection efficiencies 2 - 3.5 orders of magnitude greater than the detection efficiency of a MCP detector. The red and blue line show model values for multi-channel plates based on extensive experimental studies \cite{Liu2014}.}
\end{figure}

To solve this question quantitatively we here relate the single molecule count rate of a single pixel of $D_2$ to the electrical current that is measured by a large Faraday plate that can be positioned to intercept the ion beam at the same location as detector $D_2$. The detection efficiency was measured for singly charged vitamin $B_{12}$, $insulin^{5+}$ and $insulin^{3+}$ for two different impact velocities (energies). Mixing the molecular species and their charge states is justified since our measurements have shown that the detection efficiency should be mass independent and low-mass ions are favored here because of their higher flux, which facilitates signal comparison with a classical femto-ammeter. 
In order to normalize the current induced by a 3\,mm ion beam on a 1\,cm Faraday plate to the SSPD signal on a 200 µm detector, we measure the ion beam profile and relate the count rate to the beam shape and detector area, as shown in Figure 8 (see Suppl. Inf.). 

We measure a molecule detection quantum yield of $\eta \simeq 1$ within an energy range of 100 eV to 1000 eV. This remarkable value is 2-3.5 orders of magnitude greater than the secondary electron yield in a typical multichannel plate for ions of the same impact velocity \cite{Liu2014}. For a detailed discussion of the systematic and statistical errors, see the supplementary information. 

We find a similar area-normalized quantum yield also for of our small detectors $D_1$. An array of $D_1$ can, however, outperform an array of $D_2$ in dynamic range because a smaller pixel will only saturate at a higher molecular flux. Its smaller kinetic inductance also leads to a faster relaxation time. 
Both detector types $D_1$ and $D_2$ are clearly sufficient for most typical beams in mass spectrometry and analytical molecular beam research. The detector pixels $D_2$ are well adapted to dilute, wide molecular beams both because of their larger area and low dark count rate. Detector $D_1$ is an interesting alternative for applications with very intense molecular beams in need of high temporal resolution.

\subsection*{Cryogenic amplifiers towards scalable devices}
One would want to substantially upscale the number of pixels for applications with wide particle beams, for example in molecular impact correlation experiments, or in detectors for quantum interferometry. A large area detector such as $D_2$ is driven by a high bias current of $50-70\,\mu$A and generates a sizeable voltage pulse across a  $50\,\Omega$ impedance when the SSPD becomes normal conducting. However, the better resolving detector $D_1$ is typically driven by $> 6\,\mu$A and an impact generates only $300\,\mu$V. Low-noise cryogenic amplifiers (LNA) are thus useful to boost the signal-to-noise ratio and to reduce the rise time of the output pulse of the SSDP for later on-board digitization and time-tagging. For that purpose, we have realized an amplifier with high gain, high bandwidth, low noise, and low power consumption. We have realized such devices using a heterojunction-bipolar transistor (HBT) BiCMOS process. 

\begin{figure}[H]
\centering
\includegraphics[height=7cm]{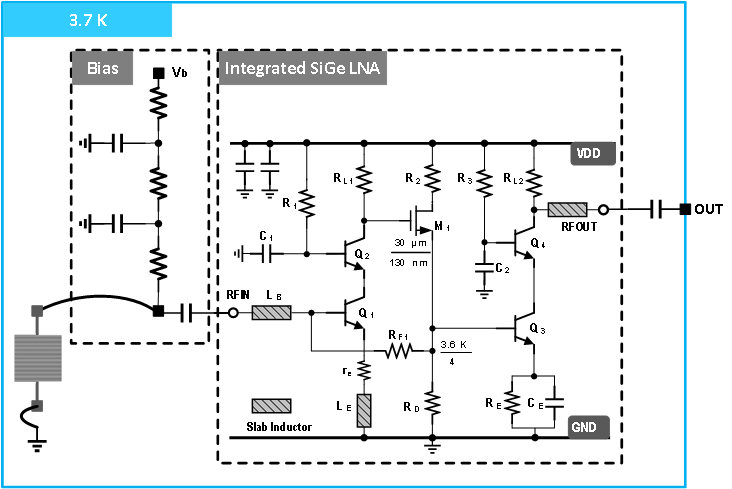}
\caption{Onboard cryogenic electronics. The diagram illustrates the integration of a SiGe low-noise amplifier (LNA) with the superconducting nanowire detector on a shared printed circuit board at $T = 3.8$\,K. These detectors with onboard electronics are used to integrate larger pixel numbers, to go to larger detection areas comparable to those of MCPs.}
\end{figure}

It exhibits good noise and speed already at room temperature and we find even an improvement of the current gain by about 300 \% at $T = 3.8\,K$ (15, 41, 42). Our device measures only $480 \times 280 \, \mu$m$^2$ including the pads and it combines an average noise temperature of $5\,$K in the frequency range of $0.1-8.8\,$GHz with a good weak-signal linearity. 
To demonstrate our integrated concept, we have joined an array of eight SSPD pixels of type $D_1$ with eight of our new LNAs to realize eight detectors $D_{1b}$ as illustrated in Figure 9 for one pixel. This system was also successfully used in ESI-QMS spectrometry on insulin, and it will allow up-scaling to much larger pixel numbers in the future. 

\section*{Discussion}
We have demonstrated that superconducting nanowire arrays can serve as highly efficient detectors for massive ions even at energies where conventional detectors fail. We can record mass spectra of singly charged insulin down to kinetic energies nearly 100 times lower than typically required for time-of-flight mass spectrometry. This is compatible and beneficial for continuous mass spectrometers, such as quadrupole or sector field instruments, where resolution improves with decreasing ion energy.

The energy dependence of the SSPD threshold current $I_{th}$ makes distinguishing n-fold charged polymers of a given mass possible, where conventional detection methods can only infer the presence of isobaric species of different charge by analysis of the isotope distribution. This can be best observed at low ion energy, where the rate of increase for the threshold current $I_{th}$ is greatest.
We have demonstrated a quantum yield close to $100\,$\% for molecules with kinetic energies of $100\,$eV or higher, independent of their structure, mass, charge or momentum. 

At lower energy the yield may be influenced by effects of mass and complexity of molecules. Molecular imaging and molecular dynamics simulations show that these entities can deform upon surface impact (43, 44). Even though vibrational excitations will eventually relax their energy to the cold nanowire this process occurs on a time scale of nanoseconds, i.e., much slower than the relaxation time of the superconducting detector. Moreover, fragmentation at high impact energy, will lead to an energy loss that might reduce the detection efficiency of the nanowire. Conversion of kinetic to internal molecular energy may thus reduce the detection yield. If molecules bounce off rather than stick to the surface, additional detection losses might occur for the detection with a SSPD, if not enough energy is transferred to the superconductor. Furthermore, the coating of the detector with a protein beam could act as a buffer for incoming particles. However, to deposit a single monolayer on the detector surface, it would need to continuously run for a few hundreds of hours in nominal operating conditions.  

While mass spectrometry and neutral molecular beam research could be served by a 10 $\times$ 10 array of detector type $D_2$, even more versatile applications with even better spatial resolution will profit from up-scaling the pixel number and using detector array $D_1$. Here we analyze some thermal limits related to on-board current sourcing, signal amplification, readout, and cabling. Even though superconducting nanowires transport current without loss, the bias current $I_b$ must be generated by sending a voltage into a resistor. For detectors of type $D_1$ ($A_1 = 400~ \mu m^2$), the power per pixel can be individually set and limited to P = 6 $\mu$W if we assume a bias current of $I_b \simeq 6 ~\mu$A, and a source resistance R = 1 k$\Omega$. Onboard current sourcing is thus still thermally compatible with 100 000 individually optimized pixels, i.e., an area of $40~mm^2$, assuming a standard 1W closed cycle cooler.  

Our experiments with detector type $D_2$ have shown that the useful sensor area can be increased by a factor of 100 while their bias current only needs to be increased by a factor of ten. The same cooling power budget would thus allow sourcing all these detectors individually across a total detector area of 400 $mm^2$. New developments in the fabrication process, may improve the uniformity of the meander, that makes the detection properties of each pixel more reproducible. This will allow to source them all with the same bias current as one large area multi-pixel array. In this case bias current sourcing no longer limits the 4 K heat budget, and the detector area can be substantially increased.

If signal amplification is required, this consumes extra power. To tackle this challenge, we have developed onboard cryogenic low-noise amplifiers in SiGe and integrated them successfully in a QMS Mass Spectrometer. They consume less than 5 mW per pixel, which allows realizing up to 200 individually amplified pixels using analog electronics. A multiplexer at 1 GHz clock frequency could sample more than 100 amplified signals, pushing the detector size again to more than 20’000 pixels using analog electronics. Such large and well-resolving devices will become important for atomic and molecular beam research with low energy particles as well as for continuous mass spectrometry with lowly charged high mass molecules.

\section*{Methods}

\subsection*{Customized mass spectrometry}
A Waters Q-TOF Ultima tandem mass spectrometer was customized as follows: a) the RF electronics of the quadrupole mass filter was upgraded by MS VISION to allow operating it up to a molecular mass to charge ratio of m/z = 30 kTh. b) a differentially pumped vacuum chamber was inserted between the quadrupole mass filter and the time-of-flight mass spectrometer to add a static quadrupole bender, static ring ion guides as well as windows for optical access. c) The ring electrodes guide the protein ions into a differentially pumped cryogenically cold ultra-high vacuum chamber, that hosts the nanowire detector arrays as well as a conventional Faraday plate. The entire cryostat is mounted on a motorized translation stage that allows to move the SSPD with 30 $\mu m$ accuracy in one axis. The detectors are cooled by a Sumitomo pulse tube cooler (RH), which achieves a cooling power of 900 mW at 4.2 K. The system is usually operated at a minimal temperature of 3.7 K compatible with all inputs and cabling, d) The mass spectrometer entrance region has also been modified to allow for protein charge reduction in bipolar air, as generated by a corona discharge.

\subsection*{SSPD working point}
In photonics, the bias current is typically set to $I_b > 0.8 ~I_c$ to maximize the quantum yield and sensitivity to low-energy photons. For molecules we typically chose $I_b \simeq 0.5~ I_c$ in $D_1$ to suppress all background contamination by photons. To avoid latching, the detector $D_2$ is written as a single wire, rather than as a device of concatenated parallel wires \cite{Cristiano2015,Ejrnaes2009,Cristiano2012}. For detector $D_2$, local amplifiers are no longer needed because the 500 nm wide nanowire is driven with a bias current of $I_b \simeq 40 - 60 ~\mu$A rather than by 4 - 6 $\mu A$ for $D_1$. A detection event triggers a peak nearly 10 times higher than for $D_1$.  

\subsection*{Cryogenic low-noise amplifier}
A single bias is used to ensure the scalability of the LNA design. Moreover, no large inductors are used for bandwidth extension. Alternatively, the capacitive peaking technique is used at the output buffer, making the LNA design very compact ($480~ \mu m \times 280~ \mu m$ including the pads).
The cryogenic BiCMOS LNA achieves an average noise temperature of about $5\,$K in the frequency range of $0.1 \sim 8.8$ GHz with well-matched input impedance ($S11 < -10 ~dB$) and sufficient gain ($> 33 ~dB$), thanks to the current resistive feedback technique and advanced HBTs used in the chip. At the expense of the power consumption, the present design obtains good weak signal linearity performance. Given that the signal level at the input of the LNA is small, the linearity can be traded with power consumption.

\subsection*{Calibration of the detector deflection electrode}	
The cold detector assembly can be vertically shifted with 30 µm accuracy and any shift induced on the ion beam by the vertical deflection electrode can thus be compensated and calibrated by a corresponding mechanical shift of the detector. Since the electrodes are symmetric in the X- and Y direction, the same calibration can be applied to derive the shift distance as a function of applied voltage. With this method one can determine an ion beam profile and deduce the absolute detection efficiency of the SSPD. (see Suppl. Inf.)

\subsection*{Acknowledgments}
We acknowledge contributions by K. Simonović and T. Sousa in parts of the mass spectrometer setup.

\subsection*{Funding}  
This project has received funding from the European Union’s Horizon 2020 research and innovation program under grant agreement No. 860713 as well as partial support by the Gordon \& Betty Moore Foundation within project \#10771. 

\subsection*{Author contributions}  
Mass Spectrometer Design and Realization: MS, PG, MFXM, AS, SD, JS, MA\\
SSPD Design \& Fabrication: NK, RG, MG, MC, AF\\
Cryogenic Electronics Design \& Fabrication: JB, CB, EC\\
SSPD based mass spectrometry: MS, MFXM, AS, SD, PG, TK\\
Supervision: MM, VK, JC, MC, AF, EC, CB, MA \\
Writing: MA, MS wiht input by all authors.

\subsection*{Competing interests} 
Single Quantum is an enterprise located in Delft/NL, specialized in the manufacturing of superconducting nanowires.  MS Vision is a mass spectrometry company in Almere/NL.

\subsection*{Data and materials availability} 	
All data, code, and materials used in the analyses will be available from the authors on request.

\newpage
\section*{Supplementary Information}
\setcounter{figure}{0}
\renewcommand{\thefigure}{S\arabic{figure}} 
\renewcommand{\thetable}{S\arabic{table}} 
\subsection*{Sample preparation \& source parameters}
\begin{table}[htbp]
  \centering
  \begin{tabular}{lllllll}
    \toprule
    Molecule & R6G & Vit B$_{12}$ & INS & CytC & Mb & Con A \\
    \midrule
    MW (kDa) & 0.44 & 1.58 & 5.73 & 12.32 & 17.74 & 25.72 \\
    Conc. ($\mu$M) & 200 & 20 & 20 & 20 & 20 & 20 \\
    Purity & 99\% & 99\% & $> 25$USP& $> 95\%$ & $>95\%$ & $> 90\%$ \\
    Solvent & C$_2$H$_3$N/H$_2$O (1:1) & H$_2$O & H$_2$O & H$_2$O & H$_2$O & C$_2$H$_3$N/H$_2$O (1:1) \\
    \bottomrule
  \end{tabular}
    \caption{Molecules and chemicals used in the experiments: Rhodamine 6G (R6G), vitamin B$_{12}$ (Vit B12), insulin bovine pancreas (INS), cytochrome C from bovine heart (CytC), myoglobine from equine skeletal muscle (Mb) and concanavalin A (Con A) were used as purchased from Merck.}
\end{table}

\begin{figure}[ht]
\centering
\includegraphics[width=14 cm]{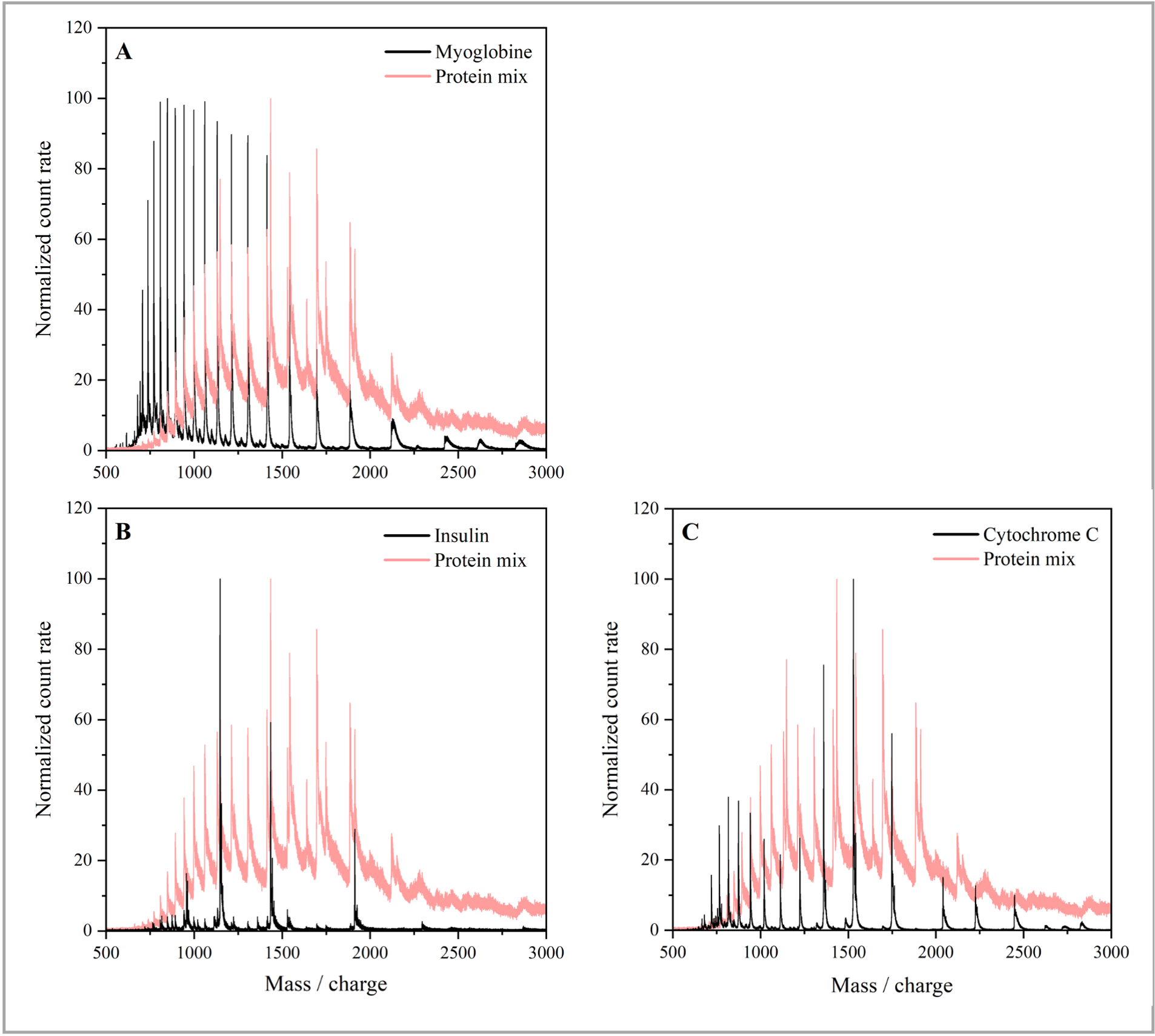}
\caption{Calibrating the mass spectrum of the protein mix.
The protein mix consisting of myoglobin (A), insulin (B) and cytochrome C (C) is electrosprayed and analyzed by TOF-MS. The peaks are identified by comparing them with the peaks for the individual proteins. The peaks are identical to those of Figure 3 in the main text.  
}
\label{MassCalibration}
\end{figure}

The molecules were  electrosprayed using a needle with an inner diameter of 80\,µm, operated at a voltage of $3-4$\,kV. The solution is driven by a Harvard syringe pump at typically 9 µl/min.	
To reduce the average molecular charge, we use a corona discharge. Nitrogen 5.0 is passed through pure ethanol at variable temperature and sent through a flow regulator valve to a chamber where a stainless steel needled at a voltage of - 2.5 to -3 kV induces the discharge. To obtain a stable ion signal the discharge current is stabilized at 15\,µA. The bipolar air is sent into an aluminum channel of 3\,mm inner diameter and 80\,mm length which is directly connected to the vacuum system. This allows to reliably reduce the charge of the proteins.

\subsection*{Microchannel plate detection}
Calibration mass spectra are obtained using the time-of-flight (TOF) spectrometer, with microchannel plate detector (MCP). The TOF spectrum is calibrated using cesium iodide. 

\subsection*{Dynode / Phosphor screen detection}
Alternatively, a system composed of conversion dynode, phosphor screen and photomultiplier detector (PD) is used to obtain an ion signal that can be directly compared to the SSPD signal. Our ion deflector works both in transmission and deflection mode, enabling the ion beam to be directed either towards the SSPD or the PD. To ensure accurate comparison, the timing is chosen such that the quadrupole completes a full scan before the deflection electrode is switched. This synchronized timing allows for a direct and reliable comparison of the spectra obtained from the SSPD and the PD.

\subsection*{Mass peak assignment for cytochrome C, insulin, and myoglobin mix}
The SSPD spectrum can be referenced to a calibrated TOF spectrum, which we show in Fig. S1. for a mixture of cytochrome C, insulin, and myoglobin and each individual compound. The peaks observed both with the MCP and SSPD, correspond to the expected compounds in Fig.\,\ref{MassCalibration}

\subsection*{Superconducting nanowire single particle detection}
The superconducting nanowire detectors are connected to a driver (Single Quantum), that sends a constant current into each pixel of the nanowire array, in the $6-100$\,µA range, and which reads out the signal across a bias-T. The signal outputs are connected to a time-to-digital converter (TDC). To initiate the measurement sequence, a programmable pulse generator (PPG) is used to provide a trigger signal for the quadrupole mass selector (QMS) scan and the TDC measurement. The PPG is necessary to synchronize the different devises needed for the measurement process. While the TDC acquires the ion count information received from the SSPD driver, the quadrupole mass selector linearly scans the RF-amplitude. This scanning process enables the acquisition of a mass spectrum using the SSPDs, as the transmission of a specific mass-to-charge state is dependent on the RF-amplitude. Since the QMS scan is linear, the SSPD spectrum can here be referenced to the calibrated TOF-MS spectrum. 

\begin{figure}[ht]
\centering
\includegraphics[height=9cm]{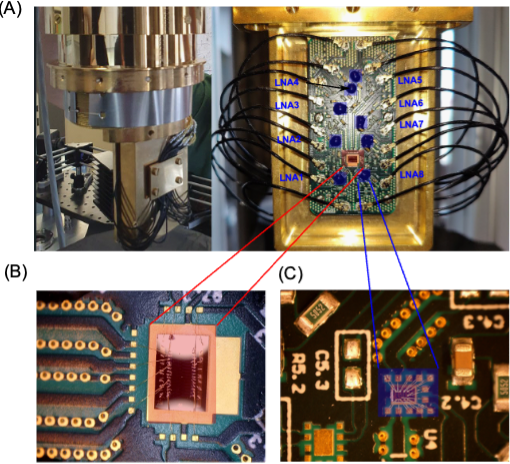}
\caption{Cryogenic low-noise amplifiers (LNAs) integrated with the SSPD chip.
Each SNWD is individually biased by a current drive at room temperature. (\textbf{A) }shows the eight-pixel readout printed circuit board (PCB) mounted onto the 3.8 K stage, as well as the biasing lines that consist of two bias lines per pixel and two supply wires for the LNAs, totaling 18 leads. \textbf{(B) } Detector array $D_1$, mounted onto the prototype PCB. \textbf{(C)} Wire-bonded LNA unit.
}
\label{cryobox}
\end{figure}

\subsection*{Cryogenic assembly}
The detector unit is shown in Fig.\,\ref{cryobox}. The detector assembly is bolted onto the cold head of a pulse tube cooler (Sumitomo SHI R65p) which can be cooled to $3.7$\,K ($P_{cool} = 900$\,mW at $4.2$\,K) with all conditioning and signal cables connected. The SSPD detector is shielded by two gold coated OFHC copper cylinders with only a 10 mm entrance hole for molecules and light. This is small enough to limit heating by black body radiation at room temperature to 36 mW. Visible photons can be shielded by covering all windows and their detection can be suppressed by choosing the bias current appropriately. 

The constant current driver is connected to the cryogenic stage via $20$\,SMA cables of $1.5$\,m length that were thermally grounded to the 80\,K, 30\,K and 3.8\,K stages of the pulse tube cooler. The full system with onboard low-noise amplifiers is shown in Fig. S1. The cryogenic assembly is mounted on a translation stage which allows us to scan the molecular beam profile and to lift the Faraday detector into the beam.  

\subsection*{Calibration of the horizontal defection electrode}
In order to calibrate the horizontal axis, a linear increase in voltage was applied to the vertical deflection electrode in steps of $10$\,V. At each voltage iteration, the superconducting nanowire detector was mechanically moved along the vertical axis.  To analyze the data and determine the effective shift as a function of voltage, a Monte Carlo method was employed. This involved fitting a Gaussian function to the measured scan multiple times, with random variations within the specified precision boundaries.

\begin{figure}[ht]
\centering
\includegraphics[width=9cm]{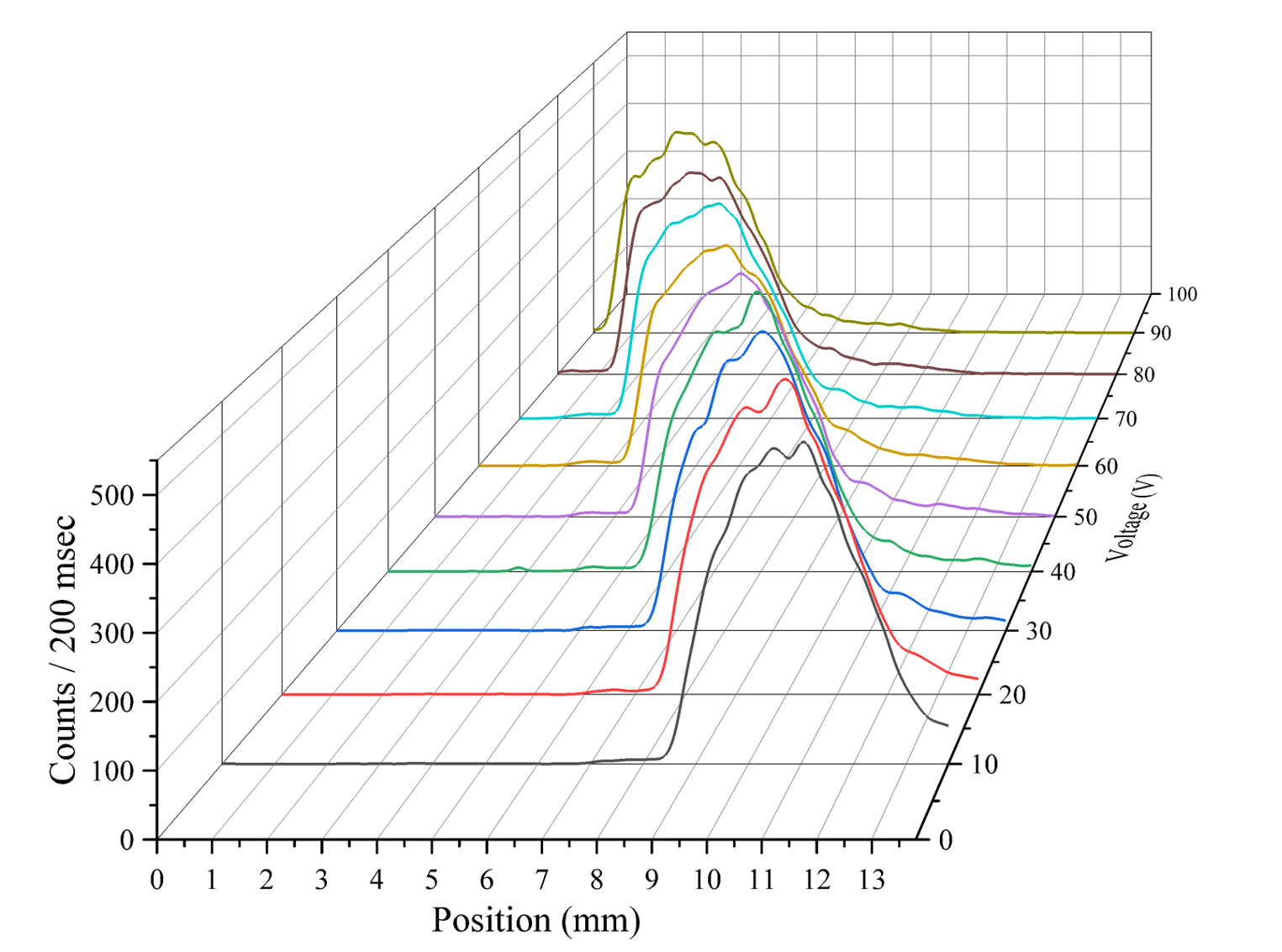}
\caption{Linear position shift of the ion beam as a function of deflector voltage, in 10V steps, for insulin at 1000\,eV. The position is scanned in 200\,µm steps by a linear stage with an accuracy of 30\,µm to monitor the deflection of the ion beam in one axis. The relevant errors can be found in Fig.\ref{DeflectionVoltage}.
}
\label{IonDeflection}
\end{figure}

\begin{figure}[ht]
\centering
\includegraphics[width=9cm]{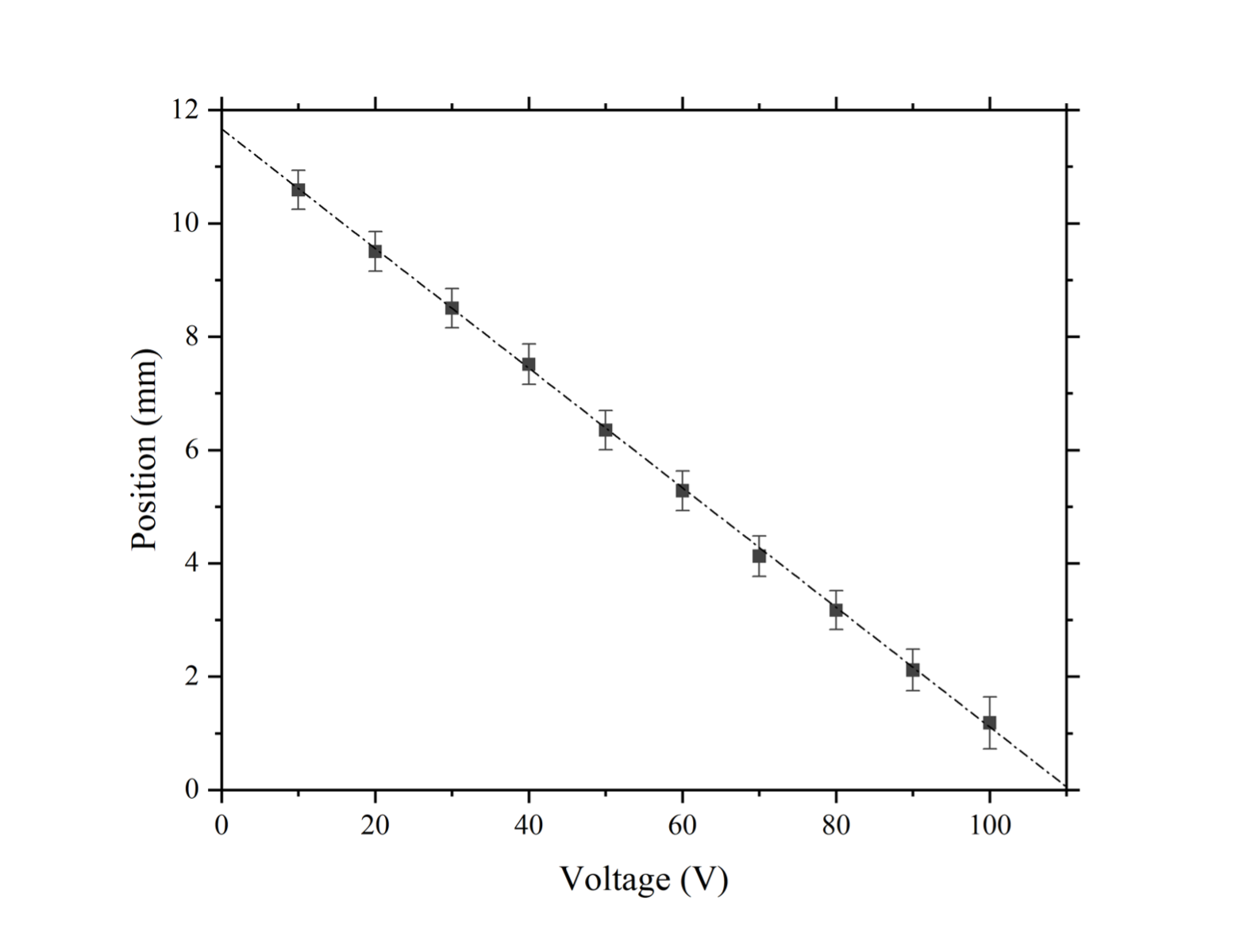}
\caption{Ion beam profilometry:
Position-shift of the ion beam at the location of the SSPD detector as a function of the deflection electrode voltage. We show the results of the Monte Carlo simulation described above.
}
\label{DeflectionVoltage}
\end{figure}

To account for the precision of the motorized linear stage, an upper bound error of $0.03$\,mm was measured. The mean and standard deviation of the center position, obtained from the Monte Carlo simulation, were plotted against the deflector voltage, along with the error due to inaccuracies in the mechanical position of the SSPD. The plot revealed a linear increase in the center position with respect to voltage, which was consistent with the original data, as can be seen in Fig.\,\ref{IonDeflection}.
To calculate the error of the slope, another Monte Carlo method was employed. The deflection voltage steps were chosen to be slightly larger than the pixel-length of the detector (200\,µm). Since the electrodes were symmetric in both the X- and Y-direction, the same calibration can be applied to derive the shift distance in the horizontal (Fig.\,\ref{DeflectionVoltage}).  
By utilizing this calibration method, it is possible to determine the ion beam profile which is important in determining the absolute SSPD detection efficiency. 

\subsection*{Systematic errors contributing to the determination of the absolute detection efficiency}
To determine the detection quantum yield we divide the incident ion count rate, as measured using the Faraday plane, by the SSPD count rate integrated over the entire ion beam profile, where we consider a meander filling factor of $50$\,\%. This way, we find a detection quantum yield close to $\eta \simeq 1$. Our observations merit some more detailed comments since Fig.\,8 suggests that the SSPD might detect even more molecules than expected by the direct current measurements, but the error bar in Fig. 8 is relevant and has several contributions: 

\begin{figure}[ht]
\centering
\includegraphics[width=9cm]{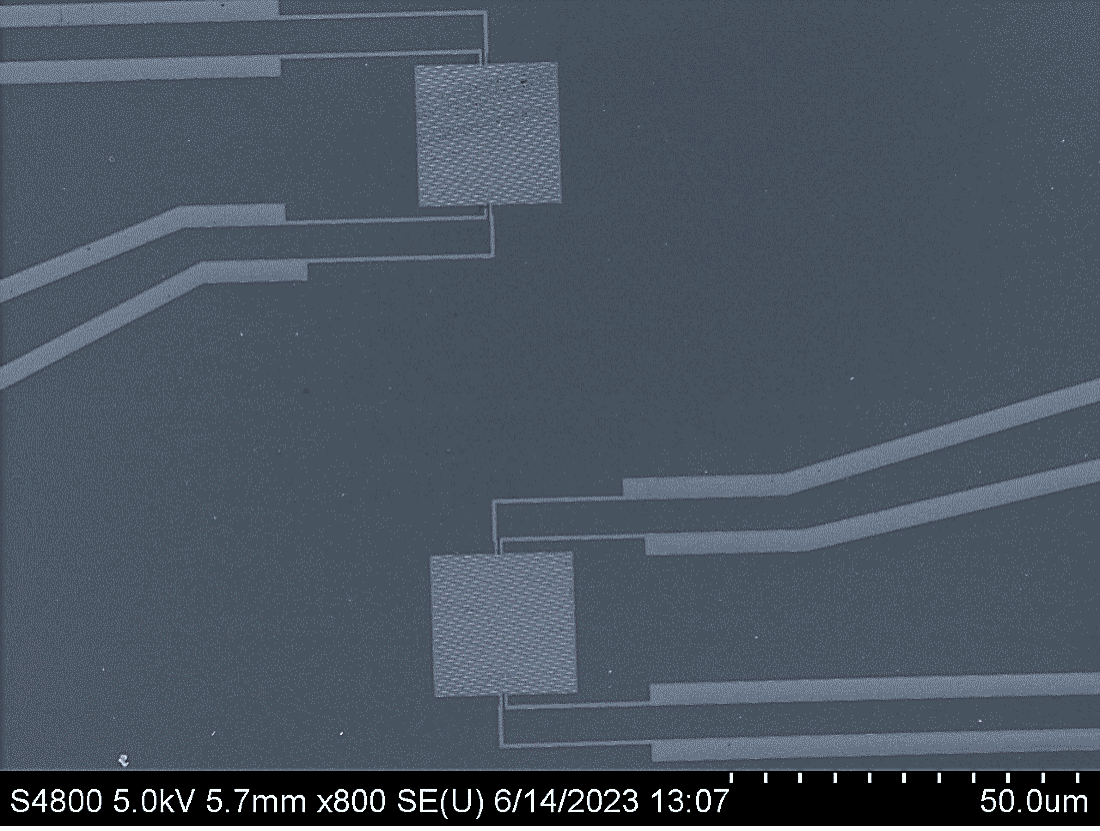}
\caption{Scanning electron micrograph of detector D1. 
Our small SSPD detector measures $20\,$µm$\times 20\,$µm and it is connected to leads which are growing in thickness with distance to the chip.  For a small detector as $D_1$, even the leads may contribute to the detection of high energy ions. For a detector like $D_2$, with an area of $200\,$µm$\times 200\,$µm this cannot contribute more than a few percent.
}
\label{D1_Leads}
\end{figure}

The protein flux is measured to vary up to 10\% over one hour. If it does the normalization will change in proportion.
	
The SSPD filling factor could deviate from the nominal value of $50$\%, and if the actual area were bigger, this could explain an excess quantum yield. However, scanning electron micrographs constrain the meander area to better than $5$\%. 

We estimate a small geometry correction factor to compensate for stitching errors that occur when sampling the $3\,$mm ion beam profile with $200~\mu$m patches of the shifted SSPD. The error of this factor amounts to $8\,$\% of the signal.

SSPD count rates would be enhanced if the ions could trigger a signal even when hitting the quartz surface in between the NbTiN nanowires. However, earlier experiments in our lab with keV atomic ions have shown that even an insulating oxide layer of only 1-2 nm thickness suffices to reduce the SSPD count rate by about an order of magnitude. We therefore estimate that heat transport across the 500 nm dielectric gaps between the meander lines contributes only on the level of 1\% to the count rate, here overestimating the yield.	 

One may argue that the protein has a finite diameter which must be convoluted with the nanowire geometry to obtain an effective detector width. For insulin or vitamin B, this contributes less than 1\% on the 500\,nm width.
	
We may need to consider the superconducting leads feeding the detector square, as shown in Fig.\,\ref{D1_Leads}. They fan out to 10 µm wide lines and can be a few 10 µm long. A $3$\,mm diameter ion beam will therefore hit some of these parts. We include the femto-ammeter, which is calibrated to better than $15$\,\%.	 

Finally, the Faraday plate can only count ions which leave at least one charge behind. If a protein hits the detector but bounces off – with deposition of energy but not charge – the SSPD can still click, while the Faraday plates will not measure any additional electric current. 

Since all these systematic and statistical errors are independent, we determine the net error bar by Gaussian error propagation which adds up to a total of $\pm 20\,$\%.

\bibliographystyle{unsrt}
\bibliography{mqo}

\end{document}